\documentclass[12pt,a4paper]{article}
\usepackage{fix-cm}
\usepackage[a4paper,left=2.0cm,right=2.0cm, bottom=2.5cm]{geometry}
\usepackage{ifthen} 
\usepackage{subfig}
\usepackage{graphicx}
\usepackage{longtable}
\usepackage[abs]{overpic}
\usepackage{multirow}
\usepackage{braket}
\usepackage{lipsum}
\usepackage{comment}
\usepackage{amsmath}
\usepackage{amssymb}
\usepackage{xspace}
\usepackage{orcidlink}
\usepackage{pgfgantt}
\usepackage{siunitx}
\usepackage{multicol}

\DeclareSIUnit\clight{\text{\ensuremath{c}}}
\DeclareSIUnit\electronvolt{e\kern-0.15ex V} 

\usepackage{booktabs}
\usepackage{lineno}

\usepackage{upgreek}
\usepackage[title]{appendix}

\usepackage{color}   
\usepackage{hyperref}
\hypersetup{linktocpage}
\hypersetup{
    colorlinks=true, 
    citecolor=blue,
    linkcolor=blue,     
}

\newboolean{pdflatex}

\newboolean{uprightparticles}
\setboolean{uprightparticles}{false} 

\newcommand*{\GeVc} {\ensuremath{\text{Ge\kern -0.1em V/}c} }

\cleardoublepage

\newcommand{\appendixnumberline}[1]{Appendix\space}

\let\oldappendix\appendix
\makeatletter
\renewcommand{\appendix}{%
  \addtocontents{toc}{\let\protect\numberline\protect\appendixnumberline}%
  \renewcommand{\@seccntformat}[1]{Appendix~\csname the##1\endcsname\quad}%
  \oldappendix
}
\makeatother

\begin{document}


\vspace*{-3.5cm}

\hspace*{-0.5cm}



\begin{center}
{\bf\Large LEP3: A High-Luminosity \boldmath $e^+ e^-$ \unboldmath  Higgs \& Electroweak \\  \vspace{1mm} Factory in the LHC Tunnel
}

\vspace*{3.0mm}
{\large A possible back-up to the preferred option (FCC-ee and FCC-hh) for the next accelerator for CERN}

{\bf Updated on October 31, 2025 }\\
\end{center}
\begin{flushleft}
C.~Anastopoulos$^{1}$,
R.~Assmann$^{2}$,
A.H.~Ball$^{3}$,
O.~Bruning$^{4}$,
O.~Buchmueller$^{5}$,
T.~Camporesi$^{6,7}$,
P.~Collier$^{4}$,
J~Dainton$^{8,9}$,
G.~Davies$^{5}$,
J.R.~Ellis$^{4,10}$,
P.~Ferrari$^{11,12}$,
B.~Goddard$^{4}$,
L.~Gouskos$^{13}$,
G. Hall$^{5}$,
M.~Klute$^{14}$,
M.~Koratzinos$^{15}$,
G.~Landsberg$^{13}$,
K.~Long$^{5}$,
L.~Malgeri$^{4}$,
F.~Maltoni$^{16,17}$,
F.~Moortgat$^{4}$,
C.~Mariotti$^{18}$,
S.~Myers$^{4}$,
J.A.~Osborne$^{4}$,
M.~Pierini$^{4}$,
P~Raimondi$^{4}$,
D.R.~Tovey$^{1}$,
D.~Treille$^{4}$,
T.S.~Virdee$^{5}$,
N.~Wardle$^{5}$,
M.~Zanetti$^{19}$
\\
\textit{         Contact person: T. Virdee (t.virdee@imperial.ac.uk)}
\vspace{-0.2cm}
\begin{multicols}{2}
\begin{footnotesize}
\textit{
$^{1}$ University of Sheffield, UK\linebreak
$^{2}$ GSI  Darmstadt, Germany\linebreak
$^{3}$RAL, STFC, UK\linebreak
$^{4}$CERN\linebreak
$^{5}$Imperial College London, UK\linebreak
$^{6}$LIP, Portugal\linebreak
$^{7}$ Boston University, USA\linebreak
$^{8}$Cockcroft Inst., Daresbury Lab. UK \linebreak
$^{9}$ University of Liverpool, UK\linebreak
$^{10}$Kings College London, UK\linebreak
$^{11}$Nikhef,  Amsterdam, Netherlands\linebreak
$^{12}$ Radboud University/Nikhef, Nijmegen, Netherlands\linebreak
$^{13}$Brown University, USA\linebreak
$^{14}$Karlsruhe Institute of Technology, Germany\linebreak
$^{15}$PSI, Switzerland\linebreak
$^{16}$Louvain University, Belgium\linebreak
$^{17}$University of Bologna and  INFN Bologna, Italy\linebreak
$^{18}$INFN Torino, Italy\linebreak
$^{19}$University of Padua and  INFN Padova, Italy\linebreak}
\end{footnotesize}
\end{multicols}
\end{flushleft}
Abstract 

The 2020 European Strategy for Particle Physics (ESPP) emphasized the critical importance of completing the High-Luminosity LHC (HL-LHC) upgrade of both the accelerator and experiments in a timely manner, identifying it as a top priority for the field. The strategy also established two key recommendations for future accelerator initiatives: (i) the realization of an electron–positron Higgs factory as the highest-priority next collider, and (ii) the investigation, in collaboration with international partners, of the technical and financial feasibility of a hadron collider at CERN with a centre-of-mass energy of at least 100 TeV, potentially preceded by an electron–positron Higgs and electroweak factory. In alignment with these objectives, the Future Circular Collider (FCC) programme—comprising FCC-ee and FCC-hh—represents the preferred path forward for CERN, offering both precision and energy-frontier capabilities. However, the 2025 ESPP update calls for the identification of prioritized alternative options should the preferred FCC pathway prove infeasible or non-competitive.
In this context, we propose LEP3, an electron–positron collider reusing the existing LHC tunnel, as a strategic backup to FCC-ee. LEP3 would exploit much of the research and development already carried out for FCC-ee, enabling high-precision studies of the Z, W, and Higgs bosons below the top–antitop production threshold. Combining strong physics potential with reduced cost, LEP3 provides performance comparable or superior to other fallback options—such as linear, muon, or LHeC colliders—while maintaining the technological continuity essential for a future energy-frontier collider. Conceived as a contingency, LEP3 complements, rather than competes with, the FCC-ee proposal.


\clearpage

\thispagestyle{empty}

\begingroup\baselineskip.99\baselineskip
\setcounter{tocdepth}{3}
\endgroup
\clearpage
\newpage 

\pagenumbering{arabic}
\setcounter{page}{1}
\section{Introduction }
\label{sec:introduction}

We discuss here the option of an e$^+$e$^-$ Higgs and electroweak factory in the LEP/LHC tunnel – LEP3 - as a possible backup for FCC if the latter is unfeasible for technical or financial reasons. This was previously proposed for the 2013 ESPP but not pursued further~\cite{Blondel:2012ey}. We follow the principal lines developed in that proposal, though the details benefit from studies for FCC-ee~\cite{2780507}~\cite{FCC_feasibility_2025_2} and incorporate many components developed for it. Thus R\&D carried out for FCC-ee would be maximally utilised. 

To maintain a high instantaneous luminosity separate rings are required for the collider and the full energy booster (accelerator), the latter for top-up injection. 

Electrons and positrons in the collider ring travel in separate beam pipes. The instantaneous luminosity is limited by the requirement of a maximum of 50 $MW$ of synchrotron radiation power loss per beam. 

The baseline anticipates two experiments. During the running programme, around $\approx4\times10^5 \, e^+e^-$ ZH events would be recorded over six years at $\sqrt{s} =230\,GeV$. LEP3 would also run at the Z and WW thresholds, recording $\approx2\times10^{12}$ Z decays over six years at or around $\sqrt{s} =91.2\,GeV$, and $\approx4\times10^7$ WW events at or around $\sqrt{s} =160\,GeV$ over four years. An alternative, shorter, run-plan could comprise 5 years of ZH running, 1 year of WW and 4 years of Z running, with a stop of 2 years after the WW running, especially if higher luminosities are attainable.

Since LEP3 would be installed in the existing LHC/LEP tunnel, it is not able to operate at the $t‑\bar{t}$ threshold. 

Using the FCC-ee cost methodology, the cost to CERN of LEP3, including costs related to civil engineering, LHC removal, LEP3 installation and two new experiments is estimated to be around 3.8 BCHF for all stages. Some 0.7 BCHF from non-CERN contributions is not included in the above number (see Table 3). 
In the costs presented here we have included the cost of two new experiments at $\approx$\ 0.35 BCHF each, as assumed for the FCC-ee experiments, with CERN's contribution amounting to 10\% of 0.7 BCHF. If the cost to the community has to be limited, the existing ATLAS and CMS experiments could be re-deployed, suitably modified/upgraded. 

We consider that the prospective physics performance of LEP3 is fully competitive with other proposed alternatives to the preferred FCC option, at a lower cost. However, LEP3 is not competitive with FCC-ee.

It is desirable that any back-up to FCC should present fewer technical and/or financial difficulties than those associated with this preferred option, as well as have good physics potential. To minimise the possibility of such difficulties, the LEP3 strategy is to re-use, as much as possible, the existing infrastructure of CERN, utilise maximally the R\&D already carried out, and keep the required financing within the envelope of the pluriannual budget of CERN. 

Some questions about the LEP3 option would need to be addressed if it is to be taken further.

\section{The Main Stages of LEP3}
\label{sec:stages_LEP3}
The LEP3 physics programme would have three phases, that should include a couple of long shutdowns, 
\begin{itemize}
    \item near or on the Z peak (91.2 $GeV$), over six years,
    \item 	near the WW threshold (163 $GeV$), over four years and
    \item near the ZH threshold (at 230 $GeV$), over six years.
\end{itemize}
The broad aim is to conduct precision electroweak physics involving the W and Z bosons as well as precision Higgs boson physics. The order proposed above corresponds to increasing the energy monotonically, but another ordering could also be considered. Running at increasing energy steps has the advantage of spreading out the required expenditure.
The duty cycle assumed is the same as for FCC-ee, i.e., 185 days @ 75\% efficiency giving an effective running time of $1.2\times10^7$ seconds per year. 

\section{Technical Considerations}
\label{tech}
The LEP3 design follows closely that outlined in \cite{Blondel:2012ey} and~\cite{2780507}. It incorporates many components being developed for FCC-ee, including magnets, RF, beam instrumentation, machine-detector interface,  etc.. The technology development plan outlined for FCC-ee would therefore be applicable to LEP3. LEP3 will be housed in the existing LEP/LHC tunnel, for which all the authorizations and permits should be in place.\\

\textit {Instantaneous Luminosity and CoM Energy for ZH Operation:}  The maximum CoM energy has been chosen to be 230 $GeV$. Although the total Higgs boson cross section peaks at a CoM energy of 250 $GeV$, the luminosity in a circular collider, at a fixed synchrotron radiation power loss, drops rapidly with beam energy, moving the peak rate of Higgs bosons lower (to 235 $GeV$ in the case of LEP3/FCC-ee).

We base our parameters on a study made by K Oide and D. Shatilov in 2017~\cite{OideShatilov} using a preliminary lattice for LEP3 inspired by the FCC-ee design. 
The Oide/Shatilov study was carried out for four IPs and at $\sqrt{s}=240\,GeV$. Their estimate of the instantaneous luminosity/IP was $1.1 \times10^{34} cm^{-2}s^{-1}$ at $\sqrt{s}=240\,GeV$ and $52 \times10^{34} cm^{-2}s^{-1}/IP$ around the Z-pole. The LEP3 baseline anticipates two IPs and ZH running at $\sqrt{s}=230\,GeV$, leading to an instantaneous luminosity/IP of $1.5 \times10^{34} cm^{-2}s^{-1}$ for ZH running. This increase, from 1.1 to 1.5$\times10^{34} cm^{-2}s^{-1}/IP$, is due to two factors of 1.15, corresponding to two experiments and a lowered $\sqrt{s}$ energy, respectively. 
The Oide/Shatilov study used very strong normal quadrupoles and sextupoles, which would lead to very high power consumption and would need a range of hardware site modifications. For these reasons we propose the use of nested superconducting quadrupoles/sextupoles. This possibility is an option in the FCC-ee design~\cite{2780507},~\cite{FCC_feasibility_2025_2}. This would result in not only a significantly lower power consumption, but also an increase in instantaneous luminosity as the magnetic bending radius could be 7 \% higher (2958m instead of 2755m), reducing the RF required as well as increasing the luminosity by 7\%. We have re-evaluated the LEP3 parameters in the light of this and the Oide/Shatilov study. The parameters from our study are shown in Table 1. 

Further work has been carried out on estimating the instantaneous luminosity using alternative parameters. The estimate given above uses a preliminary lattice developed from the one used for FCC-ee and the extracted numbers are suitably rescaled. The luminosity depends linearly on the maximum achievable vertical beam-beam parameter, which for the case of FCC-ee is 0.134. For LEP3 we have assumed 0.100. The total tune shift in FCC-ee for the case of four interaction points is 0.536, whereas for LEP3 with 2 interaction points it is only 0.200. We can therefore increase the LEP3 beam-beam parameter per IP to values at least as high as FCC-ee and most probably a factor of $\sqrt{2}$ higher, in which case the LEP3 luminosity at 115 GeV per beam will become $3.7 \times10^{34} cm^{-2}s^{-1}$. We estimate that the uncertainty in the instantaneous luminosity at each operating energy is roughly a factor of $\pm 2$. 

Another area of improvement, which could considerably decrease the cost of the project, is reducing the beam pipe diameter. The beam pipe diameter is chosen so that beam instabilities are avoided. These instabilities are proportional to the resistive wall impedance of the beam pipe, which grows linearly with the circumference of the collider and inversely with the third root of the bean pipe diameter. Therefore, to keep the same impedance as FCC-ee we can safely reduce the beam pipe diameter by a factor 1.5, going down to 40 $mm$. The cost of the magnets goes with the square of the beam pipe diameter, so the cost saving will be substantial.

We estimate that the maximum RF voltage to be installed will be around 6 GV and the instantaneous luminosity/IP at $\sqrt{s}=230\,GeV$ to be $1.6 \times10^{34} cm^{-2}s^{-1}$. For the Z running, we have changed parameters to respect two aspects: the spin modulation index, a figure of merit for spin depolarization measurements, and the threshold of x-z instability (given by $Q_s/\xi_x$ ). For the former, the problem was that the momentum compaction factor is much smaller compared to LEP (and similar to FCC). To compensate for that, we have increased $Q_s $ by increasing RF voltage. Then, to keep a reasonable beam-beam vertical tune shift, we increased the vertical emittance and decreased the number of bunches. This in turn helped in increasing the bunch length (which is good for x-z instability, resistive wall power, etc.)

\begin{table}[h] 

\hspace{0.7cm}
\includegraphics[width=0.9\columnwidth]{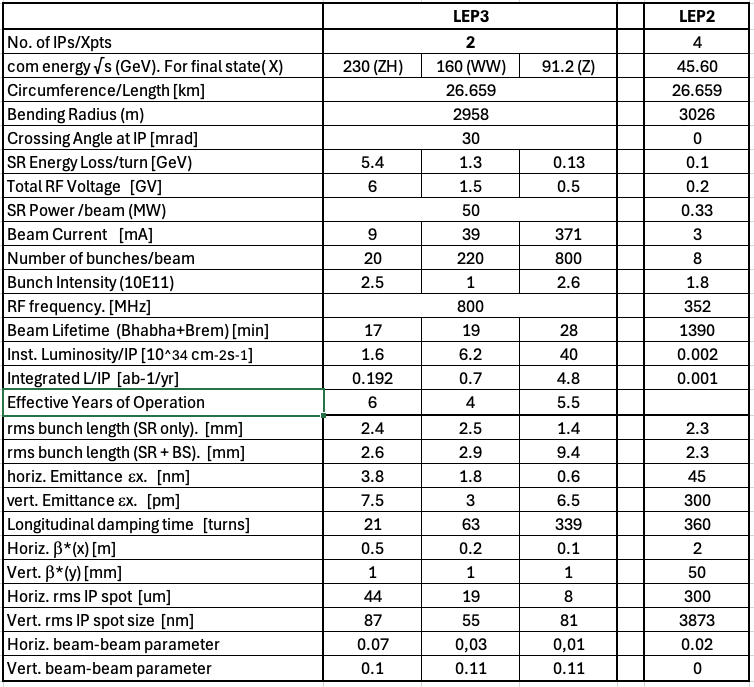}
\caption{Accelerator parameters for LEP3 }

\label{tableLEP3}
\end{table}

To calculate the number of events for the physics analyses we have used a luminosity value of  $1.6\times10^{34} cm^{-2}s^{-1}/IP$ at $\sqrt{s}=230\,GeV$, $6.2\times10^{34} cm^{-2}s^{-1}/IP$ at or around $\sqrt{s}=160\,GeV$, $40\times10^{34} cm^{-2}s^{-1}/IP$ at or around $\sqrt{s}=91.2\,GeV$. 

Hence, for two experiments, around $4.2 \times 10^5  e^+e^-  \rightarrow ZH $ events would be recorded over six years at $\sqrt{s}=230\,GeV,  1.7 \times 10^{12} Z$  decays over six years (with the first year of running at 50\% efficiency) at or around $\sqrt{s}=91.2 GeV $, and  $3.6 \times 10^7 WW $ events at or around $\sqrt{s}=160\,GeV$ over four years.  Whereas the rates of Higgs boson production at linear colliders would be similar to that potentially achievable at LEP3 for a single IP, the luminosities are considerably smaller near the Z peak and the WW threshold. Consequently, LEP3 has a much stronger precision electroweak programme, complementing its precision Higgs physics one.

\textit {Energy Calibration:} It is important for precision electroweak physics to measure the beam energy as precisely as possible for Z and WW running. The contribution of the beam energy uncertainty on the W mass measurement uncertainty at LEP, compared with that of the Z mass mesurement, was a factor 10 higher (25 $MeV$) as no resonant depolarisation measurements were possible at the WW energies (no polarisation was detected). For LEP3, we can use the FCC-ee approach for Z running, based on the knowledge that LEP routinely achieved adequate polarisation levels at the Z. For WW running where no polarization was detected at LEP, precise simulations will be needed to see if adequate levels of polarization can be achieved at energies close to or at the WW threshold. The benefits available with modern technologies, that were not available for LEP1 and LEP2, need to be investigated in addition. For example, much lower than 5\% polarisation levels might be sufficient for resonant depolarisation with today's technologies. Use of modern diagnostics and electronics will thus likely enable energy calibration at higher energies. Failing this, the aim is to use the resonant depolarisation method at the highest energy possible, and extrapolate from there to the WW threshold. We also propose to install a large amount of RF voltage for the WW running (3GV compared to an energy loss per turn of 1.5 $GeV$. This area needs further study. 

\textit {Crossing Angle and Implications: }A ``crab-waist’’ scheme with a significant crossing angle of around 30 $mrad$ (half-angle of 15 $mrad$) is necessary to attain high instantaneous luminosity~\cite{Raimondi:2007vi}. Such crossing angles necessitate special optics for the paths of the beams in the vicinity of the IPs. A scheme similar to that deployed for CEPC~\cite{CEPCStudyGroup:2018rmc} can be found for LEP3, in which the IP can be kept at the coordinates of the current IPs in ATLAS and CMS, and limiting the spatial excursion of the beams in the straight sections. We have initiated a study with experts to realise this possibility.

\textit {Placement of the Booster: }As in the case of FCC-ee, it is assumed that the full-energy booster will be installed above the two collider rings in the same tunnel. 

\textit {Booster Bypass: }Consideration has to be given to the booster/accelerator bypass around the experiments, possibly still remaining inside the experiment caverns, e.g. running along the balconies in the experiment caverns. 

\textit {Injector: }Arguments for a dedicated linac injector, rather than using the SPS as injector, are given in the FCC MTR~\cite{2780507}. Therefore, here we pursue only the linac option, retained for the FCC-ee. The injection energy is affected by non-uniformities in the bending magnets at the lowest magnetic field i.e. at injection energy. Although the injection energy potentially can be lowered by the ratio of the circumferences, i.e., by a factor of  3.4, we assume a smaller factor, and an injection energy of 10 $GeV$. This is to be compared with 20 $GeV$ for FCC-ee. The lower energy allows the linac length to be halved compared to what would be required for FCC-ee, while keeping a similar energy swing as the FCC. We adopt the siting study carried out for FCC-ee to locate the linac injector on the Prevessin site, but the shorter length allows for a more optimal placement minimizing transfer tunnel length. The layout can still be optimized to reduce the length of transfer tunnels or use existing tunnels. \\
\textit {Synchrotron Power Loss/Turn and RF Requirements:} The energy loss/turn for LEP3 operating at 230 $GeV$, with a magnetic bending radius of 2958m, is 5.4 $GeV$/turn which with a margin leads to a requirement of 6 $GV$ for RF to be installed, compared with 2.1 $GV$ for FCC-ee at $\sqrt{s}=240\,GeV$. 
Since the RF cost represents the largest item, the optimal choice of RF cavities and their powering is under intense study by  CERN’s RF group. Several criteria have to be met simultaneously e.g. power and voltage ratings for cryomodules (CRMs), ratings for power couplers, lateral and longitudinal integration in the LSS etc. The choices presented here have been inspired by the designs retained for the FCC-ee. We believe that the RF system can be further optimised with a view to lowering the overall cost. For WW and ZH running the CERN-RF Group proposes the use of  800 $\mathit{MHz}$, 4-cell RF cavities, similar in design to the FCC-ee 6-cell cavities. These 4-cell cavities should be shorter in length, and the cryomodules shorter by about 1.5m. Due to the high current during Z running the RF Group proposes the use of 800 $\mathit{MHz}$ single cell cavities. Further, they recommend the use of 800 MHz 6-cell cavities for the booster. Costs of installation,  removal and replacement of CRMs are included in the overall cost estimate. A summary of the RF cryomodules/cavities is shown in Table \ref{Table_RF}.

\begin{table}[h!] 

\hspace{2.5cm}
\includegraphics[width=0.7\columnwidth]{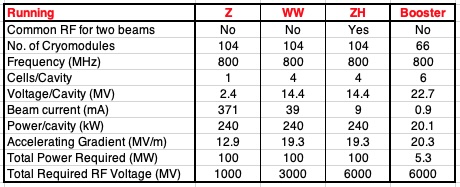}
\caption{Parameters for the RF system for Z, WW and ZH running}

\label{Table_RF}
\end{table}

A maximum power of 250 kW is assumed for each power coupler. All four even-numbered Long Straight Sections (LSS) will be used for the RF; two for the collider ring (with separators/recombinators) and two for the booster. The LSS have the same length of 545 $m$ with a diameter of 4.4 $m$. The same cavities are used for the electrons and positrons for ZH operation. Adopting the new  4-cell design for RF cryomodules, with a length of 8.45 $m$ for each cryomodule, leads to 52 cryomodules in each of two LSS. Placing one quadrupole with a length of 3.6 $m$ for every  8 cryomodules leads to $52\times 8.45+25m = 465m$. Reserving $2 \times 40\,m$ for electrostatic separators leads to a total length of 545$m$.

\textit {Civil Engineering Works:} The LEP/LHC tunnel was constructed in the 1980s with an assumed lifetime of some 40 years. To prolong its usability, either for LEP3 or as an injector for the future FCC-hh programme, or for reasons of safety, maintenance will be needed in some sections considered to be fragile. It is estimated that this repair corresponds to a section of a length of 0.6 $km$ in Sector 3-4. Furthermore, the straight sections around the two collision IPs will need widening to accommodate the large crossing angle. The length of the tunnel assumed to be widened corresponds to two times 2$\times$270 $m$, with a maximum diameter of ~7 $m$. Allowance has been made for some additional cores from the klystron galleries into the tunnel. It is assumed that the existing LHC shafts can be used to carry out the LEP3 civil engineering works. The civil engineering works are estimated to cost around 165 MCHF.

\textit {Radioprotection Issues: }Studies have been carried out in the context of HL-LHC that indicate that the activation levels inside the LHC tunnel are low (surface of the tunnel walls and machine elements inside). Discussions with CERN's RP Group suggest  that there are no showstoppers but further studies would be needed before drawing firm conclusions and on how the dismantling work should proceed. 

\textit {Experiments:} Although having two new experiments in the existing ATLAS and CMS caverns may be preferable for many reasons, including a better-adapted physics performance, also use could be made of the two existing LHC general-purpose experiments i.e. ATLAS and CMS suitably modified or upgraded, including their inner tracking systems. Which option to adopt will require an optimization and considerations concerning the cost, the physics performance and sociology. If this Project proceeds it is probable that a call for experiments would be made and new and/or suitably upgraded existing experiments inevitably would be proposed. A previous study explored the suitability of the CMS experiment for such e$^+$e$^-$ operation~\cite{Azzi:2012yn}. 
The R\&D carried out by the ALICE collaboration for its new bendable pixel detector could form a possible basis for LEP3 pixel and tracking detectors. Special luminometers will be required. If existing detectors are used, finer lateral granularity, e.g., for hadronic calorimeters may have to be introduced. The TDRs for the experiments (new or re-deployed) would have to be submitted by the mid-2030s, several years before the start of the shutdown for the transition to LEP3, enabling a head start for construction. The trackers would be designed to be light-weight, with a momentum resolution for 40 $GeV$ charged particles at the $\sim 0.1$\% level, and an angular resolution on 40 $GeV$ muons of $\sim 0.1\,mrad$.
If ATLAS is reused, modifications to the endcap toroid magnets and/or liquid argon calorimeter may be required to accommodate the final focus of the collider ring. 
Some level of particle identification can be obtained from the precision timing detectors being installed for HL-LHC running. 
However, unless the possibility of two new detectors is adopted, probably there is no room to introduce significant $\pi$/K separation capability.

We estimate that two new experiments would cost the community some 2$\times$0.35 BCHF, as in the case of experiments for FCC-ee. The cost to be borne by CERN is assumed to be 10\%, as in the case of FCC-ee costing, ie. 0.1$\times$2$\times$0.35 BCHF = 70 MCHF. This has been included in the cost estimate presented here. In existing detectors, new inner trackers would cost around 100 MCHF per experiment, and we estimate that another 50 MCHF would be needed for all other modifications.
It is assumed that a large fraction of the existing experimental collaborations would remain to exploit data from LEP3. If ATLAS and CMS are re-deployed there would be a considerable advantage in using understood detectors, software and analysis tools, and scientists already experienced in physics analysis of data from these existing experiments.
For Z running the inner detector would be immersed in a solenoid field of 2T in both the ATLAS and CMS experiments. In CMS the solenoid magnetic field could be raised for higher  $\sqrt{s}$ running, e.g., 3 Tesla during ZH running. In ATLAS it is expected that the toroid magnets for muon momentum measurement will continue to operate, possibly at a reduced field. The rest of the ATLAS and CMS detectors should continue to function well during the duration of LEP3 programme. 

\textit {Machine-Detector Interface: }There would be superconducting final quadrupoles near the interaction regions that follow the design outlined in the FCC MTR~\cite{2780507}. Starting at a distance of 2$m$ from the interaction point, a shielding solenoid surrounds the quadrupole, and a compensating solenoid is located close to the collision point. A luminometer is integrated. A stay-clear cone is defined, above 100 $mrad$, from the IP along the z axis. This corresponds to a line with pseudo rapidity of $| \eta |\sim 3$. Thus the current forward or very forward calorimeters will not be required.

\textit {Cost Estimate: }The capital cost of LEP3 has been evaluated employing the methodology used to estimate the cost of FCC-ee, with appropriate scaling (e.g., by the numbers of components) of the costs of individual FCC-ee items. The error in this estimate would be similar to that for the FCC-ee. Included in this estimate are the costs of the civil engineering works described above, the cost of careful removal and storage of LHC components to allow possible future re-use, and the installation of the LEP3 machine. The latter costs have been estimated by discussions with those involved in the installation of the original LEP accelerator. The total cost to CERN of the LEP3 project, with two new experiments, is estimated to be 3.8 BCHF (see Table \ref{table2}). The cost is dominated by the cost of the RF system ($\sim 1.5$ BCHF). The next costliest items are the quadrupole/sextupole magnets ($\sim 0.4$ BCHF) and the injector chain ($\sim 0.3$ BCHF). The FCC costing study uses a classification matrix published by the Association for the Advancement of Cost Engineering (AACE).

\begin{table}  

\hspace{3.5cm}
\includegraphics[width=0.6\columnwidth]{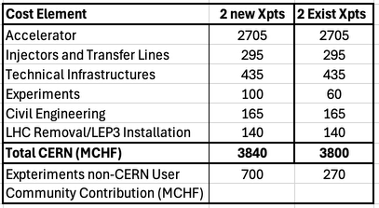}
\caption{The LEP3 cost to CERN is estimated to be 3840 MCHF using the FCC-ee costing methodology excluding the additional cost of 700 MCHF to the worldwide particle physics community. The same costs for two existing experiments amount to 3800 MCHF excluding 270 MCHF respectively. These costs are dominated by the cost of the RF system at over 1.5 BCHF}

\label{table2}
\end{table}

\textit {Schedule:}The LEP3 accelerator would be installed after the end of the HL-LHC programme. We estimate that five years will be needed for dismantling the LHC, carrying out civil engineering works and installation and commissioning of the LEP3 accelerator. This estimate has been made by some of those previously responsible for the installation, commissioning and removal of the LEP accelerator; and the installation and commissioning of the LHC accelerator. Assuming the current HL-LHC schedule, LEP3 could be operational in the second half of 2040s. 

\textit {R\&D Needs: } It is estimated that around 25 MCHF and 40 FTEy, over and above the resources foreseen for the FCC, will be required to advance the project sufficiently to reach a point of decision, anticipated in around 3 years time. The FTEy and financial resources mentioned above are foreseen for prototyping (e.g. of nested superconducting quadrupole and sextupole magnets, fundamental power couplers, etc.) or further studies (including those for RF integration in the existing tunnel, developing a low-emittance LEP3 lattice and optics, booster bypass, injector and its layout, crossing angle, radio-protection related, synchrotron radiation absorption etc.) or consolidation (e.g. of the schedule for the shutdown activities, costing  etc.).  We believe that around 5\% of the capital cost of the Project will have to be spent for the remaining R\&D and prototyping before proceeding to construction. This would amount to about 200 MCHF. In addition to the R\&D needs for the FCC-ee project outlined in submission ~\cite{FCC_feasibility_2025_2}, two particular areas of R\&D affecting LEP3 are called out above: R\&D on higher power fundamental power  couplers, and on nested superconducting quadrupole/sextupole magnets. 

\textit {Sustainability: }We assume that all building permissions/permits for LEP3 should either be in place or be easy to obtain. Also, the amount of preparatory work and the environmental impact should be minimal. As relatively little construction is required, it is expected that the equivalent carbon cost would be a fraction of any new build. Few other significant environmental impacts are anticipated. Nevertheless, benefit would be drawn from any relevant studies in these areas carried out for other proposed projects such as FCC-ee or CLIC. Usage of high-temperature superconducting arc quadrupoles and sextupoles (HTS) magnets would help limit the power consumption compared to using warm magnets. 

\textit {Power Consumption:} We estimate that the total power consumption at the highest LEP3 energy will be about 250 $MW$, and about 185 $MW$ at the Z pole. It is likely that low temperature (LTS) or high temperature (HTS) superconducting quadrupole and sextupole magnets will be needed to minimise power consumption.

\section{The Physics Case for LEP3 }
\label{physics}
The aim is to conduct precision studies of Higgs boson physics and of electroweak physics involving W and Z bosons. The precision measurements are made possible by recording large samples of W, Z and H bosons. 
\subsection{Higgs Boson Physics}
The discovery of the Higgs boson has raised several theoretical questions. These questions motivate attempts to obtain a deeper understanding of the physics of the Higgs boson and the high priority of a Higgs and electroweak factory, as recommended by the 2019 ESPP, as well as searches for direct clues to new physics. During the 2030s the study of the Higgs boson will be carried out at the upgraded HL-LHC accelerator and experiments, with the goal of studying some ten times more proton-proton collisions than originally foreseen. A question that can be posed is: what is the needed precision for the measurement of the properties of the Higgs boson? One answer is to measure these as precisely as possible. But this begs the question: what extra is gained by a precision that is factor 2-3 times better, i.e., how precise is precise enough? Nevertheless, it is clear that an e$^+$e$^-$ Higgs factory would be able to probe Nature with unequalled precision.
The precision measurements possible at LEP3 in the Higgs boson sector are compared, in Table \ref{table3}, with the situation at the end of HL-LHC. The LEP3 numbers are for two experiments. The preferred FCC-ee option would provide a better precision for a large range of Higgs boson coupling measurements.

\begin{table}[h!] 
\centering
\includegraphics[width=0.7\columnwidth]{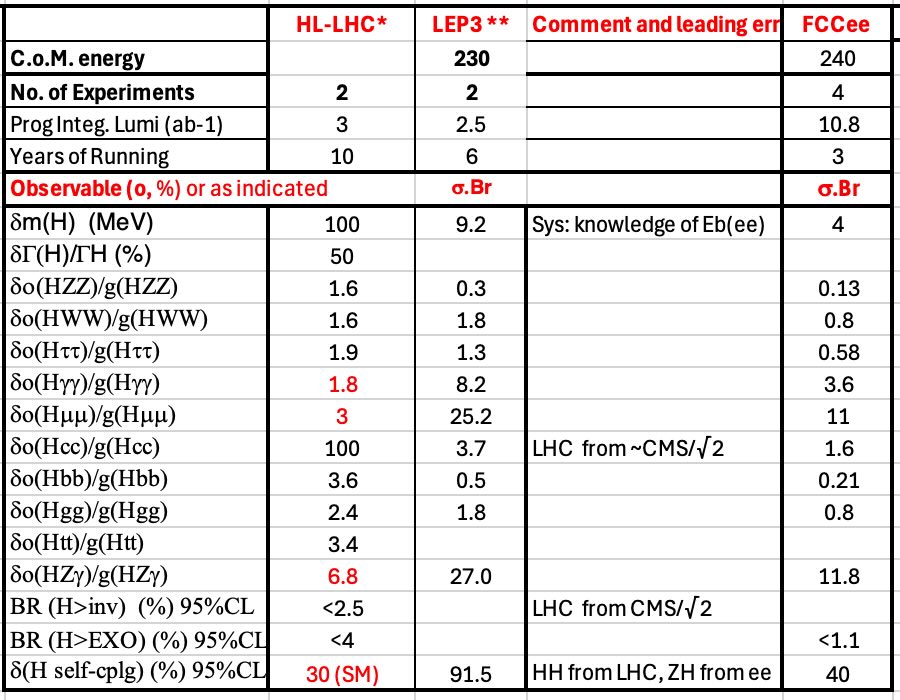}
\caption{The precision for Higgs boson couplings measurements. The errors for LEP3 have been increased with respect to the recent FCC-ee submission ~\cite{FCC_Physics-ESPP25},  by a factor $\sqrt{22/4.2}$. In red are HL-LHC values that are unlikely to be improved by $e^+e^-$  Higgs boson factories.}
\label{table3}
\end{table}

\subsection{Electroweak and Flavour Physics}
The physics case for circular $e^+$e$^-$ colliders is considerably 
enhanced by the possibility of precision electroweak 
measurements. Precise measurements at the Z peak and WW 
threshold will form an integral part of the almost 20-year LEP3 
running programme.
Heavy flavour (c, b, $\tau$) physics can be studied at the Z peak along with perturbative QCD and hadronisation, searches for rare decays including those forbidden in the Standard Model, and the exploration of neutrino mass models invoking right-handed neutrinos.

\begin{table} [h] 
\hspace{1.5cm}
\includegraphics[width=0.8\columnwidth]{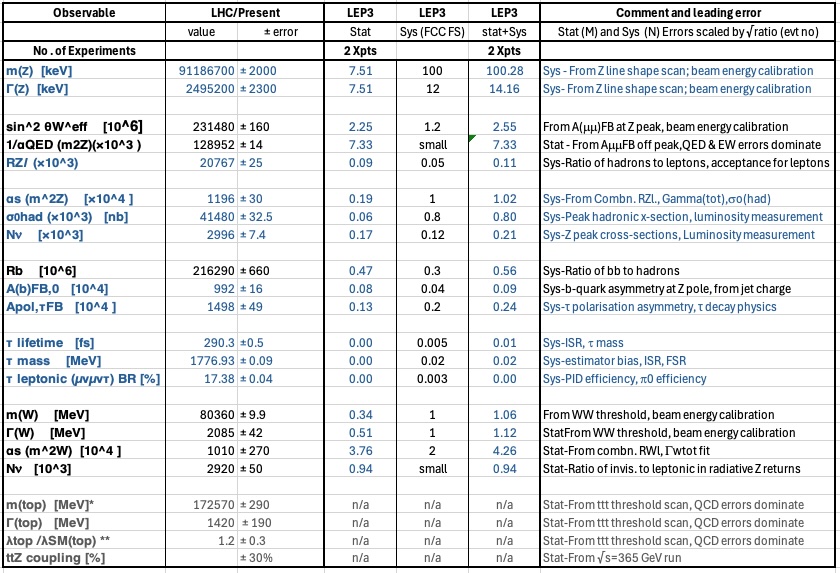}
\caption{Experimental precision (statistical + systematic) of a selection of electroweak mea-
surements at LEP3. The statistical errors are scaled by a factor $\sqrt{6/1.7}$ from the FCC-ee submission ~\cite{FCCEE-ESPP25}, ~\cite{FCC_feasibility_2025_1}, ~\cite{FCC_Physics-ESPP25} whilst the systematics ones are kept the same.}
\label{table4}
\end{table}

In addition to flavour physics and BSM searches, the electroweak measurements will include measurements of m$_Z$ , $\Gamma_Z$ , m$_W$, $\Gamma_W$, sin$^2\theta_{Weff}$, R$_b$ , $\alpha_{QED}(m_Z)$,  $\alpha_s(m_Z)$, electroweak couplings, etc., with unprecedented precision. 

We expect to record $1.7 \times 10^{12}$ visible Z boson decays, including $ 2.1 \times 10^{11} B^0/\bar{B}^0, 2.1 \times 10^{11} B^+/B^-, 5.1 \times 10^{10} B_s^0, 1.3 \times 10^9 
B_c$ decays, as well as $1.7 \times  10^{12}$ charm quark pairs and $ 6 \times  10^{10} \tau^+ \tau^- $ pairs. This will enable an extensive 
high-precision flavour physics programme at LEP3, largely complementary to the HL-LHC and Belle II programmes. 
In particular, precision measurements of fundamental parameters of tau leptons (lifetime, mass, leptonic branching fraction), 
rare $ B_s^0$ and $B_c $ decays and b baryon decays with missing energy, and charm physics will yield very rich and often unique results.

Given the much higher numbers of recorded Z and WW events, LEP3 would be considerably better than ILC(250) in making precision electroweak measurements.

As has been discussed, e.g., in~\cite{Maura:2024zxz}, precision electroweak measurements around the Z peak provide indirect sensitivity to new physics at scales tens of TeV through higher-order loop effects. These indirect effects provide probes of potential deviations from the Standard Model that are complementary to those obtainable from measurements at higher energies above the Z pole. Explicit examples of how accuracy complements energy for operators within the SMEFT framework are given in~\cite{Maura:2024zxz}. A high-precision Tera-Z programme may thus anticipate aspects of physics runs at higher energies and provide a wide scope of quantum exploration of the TeV scale. See Table \ref{table4} for the expected perfomance of LEP3.

\section{The Longer-Term Future }
\label{future} 
The first priority of the European Strategy should remain the exploitation of the full potential of the LHC, i.e., to successfully complete and fully exploit the HL-LHC upgrade of the LHC accelerator and its experiments. An obvious question is: what comes next?

A major objective in particle physics is always to operate an 
accelerator that allows a leap of an order of magnitude in the 
constituent centre-of-mass (CoM) energy with respect to the 
previous one. Following on from the LHC, which operates at 
constituent $\sqrt{s}$ $\sim$ 1 $TeV$, the two possibilities to attain 
$\sqrt{s} \sim$ 10 $TeV$ directly are the FCC-hh operating at 
$\sqrt{s}$ $\sim$ 100 $TeV$ or a muon collider operating at $\sqrt{s}  \sim$ 10 $TeV$.  

Neither of these possibilities is currently ready to proceed to full construction. Were industrial high-field (14-16 T) magnets already existing, our preference would be to go immediately to FCC-hh, giving a constituent CoM energy a factor of 7 higher than the LHC. Were a muon collider ready to be built, it would be a serious candidate for the near future, as the requirements for civil engineering and hardware would be moderate in comparison. However, an extensive programme of R\&D is required before construction of a muon collider could be approved.

Considering the preparation times needed for the above projects, another accelerator is needed to bridge the gap between the end of HL-LHC and a higher-energy collider that provides direct probes of physics at a constituent $\sqrt{s} \sim$ 10 $TeV$. 

Whatever is the accelerator covering this gap, we expect that it would have to be built within the financial resources currently available to CERN and its particle physics user community. Any proposed project should address the recommendation (c) from the 2020 ESPP, which we interpret as requiring the production of a large sample of Higgs bosons for a more powerful analysis than is currently possible. It should also be able to explore the 10 $TeV$ scale indirectly via a suite of high precision electroweak measurements.
For these reasons, we think that the LEP3 accelerator proposed here would provide an ideal back-up in the event of technical or financial issues for the preferred FCC project.

\textbf{We reiterate that we support FCC as CERN’s next major project and consider LEP3 only as a back-up option in case FCC-ee turns out not to be technically or financially feasible. }

Given its lower luminosity and inability to reach the $t‑\bar{t}$ threshold, LEP3 would be a less capable machine than FCC-ee (or CEPC). However, if neither FCC-ee nor CEPC proceed, LEP3 could be a more viable  e$^+$e$^-$ Higgs boson and electroweak factory than any of the other proposed back-up options. The question then arises: what does the far future hold for CERN? What would follow LEP3? 

A $\sim$100 $TeV$ hadron collider would be the obvious next step, for which high-field magnets are required. Hence R\&D on these must be vigourously pursued, followed by the preparation for industrial production to provide solid cost estimates. 
Muon colliders may provide an alternative approach to reach constituent CoM energies of 10 $TeV$ or higher. Hence R\&D on muon colliders must also be pursued with equal vigour. Such a programme should have milestones along the way that would deliver physics as the R\&D progresses, e.g., experiments involving muon-driven neutrino beams such as nuSTORM or neutrino/antineutrino factories.
If the future lies with a high-energy hadron collider such as FCC-hh, and if FCC-ee proceeds, then the required tunnel would already exist. However, if FCC-ee does not proceed, the FCC-hh tunnel could also be dug later, profiting from the site studies currently underway, and giving time for magnet development/industrialization. On the other hand, if it turns out that the muon collider is the preferred path, the required tunnel would be smaller than even the LEP/LHC tunnel, which could host an intermediate/injector ring.
To reach global consensus on what path to take, it is preferable that the future strategy deliberations of different regions coalesce into a single shared strategy, with all regions committing to building the commonly understood preferred option. This would facilitate gathering the necessary financial resources for the construction of FCC-hh or SppC, including the tunnel if it does not already exist, or building a muon collider at a globally acceptable site.

\section{Summary}
\label{summary}
\textbf{We support FCC as the preferred option for CERN’s future}, but recall that the guidance for the 2026 ESPP strategy review requests proposals for alternatives to this preferred option. Linear e$^+$e$^-$, circular muon and LHeC colliders are among the alternatives that have been proposed.  
\textbf{We propose that another option, an e$^+$e$^-$collider in the LHC tunnel, referred to here as LEP3, also be considered as an alternative, possibly as the primary alternative. }
We have described its capabilities as an e$^+$e$^-$ Higgs and electroweak factory. Compared to the linear e$^+$e$^-$ colliders proposed, LEP3 with two effective experiments would make measurements with similar precision in the ZH mode and with superior precision at the Z and WW energies, all at much lower cost. 
No showstoppers have yet been identified, and we consider this proposal to be sufficiently interesting to deserve further study and investment. That said, we have identified important areas that would require deeper investigation before CERN could commit to LEP3.

\clearpage
\bibliographystyle{JHEP} %
\bibliography{refs_LEP3.bib}

\end{document}